\documentclass[10pt,conference]{IEEEtran}
\IEEEoverridecommandlockouts
\usepackage{cite}
\usepackage{amsmath,amssymb,amsfonts}

\usepackage{graphicx}
\usepackage{textcomp}
\usepackage{xcolor}
\usepackage{algpseudocode}
\usepackage{algorithm}
\usepackage{subfigure,multirow}
\usepackage{url}
\usepackage{color}
\usepackage{gensymb}

\usepackage{array}
\usepackage{amsmath}
\usepackage{mathrsfs} 
\usepackage{autobreak}
\usepackage{bbm}
\usepackage[numbers]{natbib}

\def\BibTeX{{\rm B\kern-.05em{\sc i\kern-.025em b}\kern-.08em
    T\kern-.1667em\lower.7ex\hbox{E}\kern-.125emX}}
\begin{document}

\title{IntLearner: AI-enabled Interference Mitigation for Wireless Networks  \\}

\author{
	\IEEEauthorblockN{Ruirong Chen, Gaoning He}
	\IEEEauthorblockA{2012 Labs, Huawei Technologies
		Shanghai, China
		\\\{ruirongchen, hegaoning\}@huawei.com}
}

\maketitle

\begin{abstract}
The future Six-Generation (6G) envisions massive access of wireless devices in the network, leading to more  serious interference from concurrent transmissions between wireless devices in the same frequency band. Existing interference mitigation approaches takes the interference signals as Gaussian white noise, which cannot precisely estimate the non-Gaussian interference signals from other devices.  In this paper, we present IntLearner, a new interference mitigation technique that estimates and mitigates the impact of  interference signals with only physical-layer (PHY) information available in base-station (BS) and user-equipment (UE), including channel estimator and constellations. More specifically, IntLearner utilizes the power of AI to estimate the features in interference signals, and removes the interference from the interfered received signal with neural network (NN). IntLearner's NN adopts a modular NN design, which takes the domain knowledge of BS and UE PHY as the guidance to NN design for minimizing training confusion and NN complexity. Simulation results show IntLearner increases Uplink (UL) channel estimation accuracy up to 7.4x, and reduces the Downlink (DL) Signal to Interference Ratio plus Noise Ratio (SINR) requirement to achieve the same Block Error Rate (BLER) by 1.5dB in a conventional multi-cell scenario.

\end{abstract}

\begin{IEEEkeywords}
Neural network, Interference mitigation
\end{IEEEkeywords}

\section{Introduction}
Radio interference occurs with multiple concurrent wireless transmissions in the same frequency bandwidth, which distorts the transmitted signals and degrades wireless network performance. More inter-cell and inter-user interference at UEs occur with a much denser deployment of BS to support more UEs in the next 6G network enhanced mobile broadband (eMBB) \cite{eMBB2021Dog} and massive Machine Type of Communication (mMTC) \cite{MMTC5G,MMTC2021}, which leads to higher network performance loss. To enhance network connectivity and throughput in 6G, it is crucial to mitigate inference during transmission.

Traditional  Maximum Ratio Combining (MRC) \cite{MRC1999,MRC2003,MRC2009,MRC2014Tan} receiver aims to mitigate the impact of interference by taking the interference signal as Gaussian white noise, which introduces large estimation error from the difference in distribution between noise and interference. To further improve interference cancellation accuracy, Interference Rejection Combining (IRC) \cite{IRCLTE_2012,IRC2011,IRC2014liu} has been proposed to find the best receiving direction with interference for multiple-input and multiple-output (MIMO) system by estimating the direction of interference from the  difference in received signals at MIMO antennas.  However, IRC fails with a single antenna receiver, due to the inadequate direction information from multiple antennas. Other Interference mitigation approach \cite{SIC1994Pat,SIC2011li,SIC2015hig,SIC2012mir}, separates the interference and transmitted signal by removing the reconstructed  interference signal from the mixed signal, and decode the transmitted data from the residue signal. But such a scheme requires the information of interference available to the receiver, which cannot mitigate the impact of unknown interference sources.

Instead, we envision a fundamental shift in real-time interference mitigation for cellular networks: instead of taking all interference as white noise, the knowledge of interference can be learned from the PHY information \cite{PHYinfo2015poo}, such as channel estimator \cite{ChanEST2011ber,ChanEst2009hou} and constellations. Such learning is feasible due to the distinguishable difference exhibited from interfered signal between the non-interfered and interfered channel estimators. In this way, the learned interference can be removed from the distorted constellations in data symbols to recover the transmitted data without inference.
\begin{figure}
	\centering
	\includegraphics[height=1.8in]{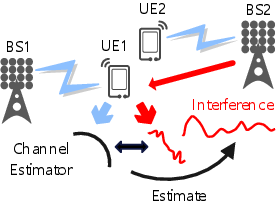}
	\vspace{-0.05in}
	\caption{Interference Estimation with PHY information}
	\vspace{-0.25in}	
	\label{fig:Gen_des}
\end{figure}

Based on such insight, we propose IntLearner, a novel real-time interference estimation and mitigation approach with PHY information available in BS and UE without hardware modifications for both UL and DL. IntLearner estimates the characteristics of interference by comparing the difference in phase and amplitude of the interfered and non-interfered channel estimators at the receiver, as shown in Figure \ref{fig:Gen_des}. Then, IntLearner applies the knowledge of interference to mitigate the impact of interference in UL channel estimation and DL data transmission by removing the interference from UL channel estimator and DL data symbols via regression.


The key challenge for IntLearner is estimating the interference and recovering the channel from channel estimators without any prior knowledge of the interference sources. For example, an interfered channel estimator could be a mixture of a channel with low Signal-to Noise Ratio (SNR) and an interference with low signal strength, or a high SNR channel with strong interference.  Our solution for this challenge is utilizing the power of Artificial Intelligence (AI) to 
precisely estimate the interference and recover the channel. To reduce the ambiguity for NN training, we design a modular network with the domain knowledge of modules in UE and BS PHY. By doing so, IntLearner enhances the training quality and estimation accurate of the NN.


More specially, the channel estimators may not accurately represents the channel at each resource element (RE) with the estimation from discrete Sounding Reference Signal (SRS) and Demodulation Reference Signal (DMRS) in the frequency domain. IntLearner extracts and refines the features of interference estimation at each RE via training a CNN for discrete channel estimations. Then the estimated interference can be removed with a fully-connected NN, mimicking the removal of interference in frequency domain. However, due to the randomness in interference, the interference may not be completely removed from the transmitted signal. IntLearner further strengthen the interference removal by exploiting the correlation between constellations from encoding with an LSTM network to recover the transmitted data.


In practice, IntLearner can be easily implemented into UE and BS PHY without any hardware modifications for enhancing the performance of cellular networks with interference.  Our detailed contributions are listed as follows:
\begin{itemize}
	\item We proposed a novel NN design to estimate and mitigate the impact of interference in both UL and DL with only PHY information available at BS and UE.
	\item Our NN enhances  7.4x UL channel estimation  accuracy with low complexity by taking channel estimation in current BS PHY design as the domain knowledge to minimize the confusion for NN.
	\item Our NN corrects bit errors and reduces 1.5dB SINR requirement to achieve the same BLER by accurately remove the signal distortion from interference in the received constellations.
	
\end{itemize}


\section{Background and motivation}
In commodity BS and UE, the PHY information, such as channel estimators, constellations and data bits, are the intermediate outputs of sub-process in PHY, which can be accessed without hardware modifications. To better understand the utilization of PHY information for IntLearner, we first introduce the background for acquiring and processing the channel estimators for UL/DL. We then motivate our design by demonstrating the feasibility of interference estimation with channel estimators, and limitation of mitigating the interference with a monolithic NN from channel estimators.
\begin{figure}[ht!]
	\centering
	\subfigure[Antenna configure 1] { 
		\includegraphics[width=1.6in]{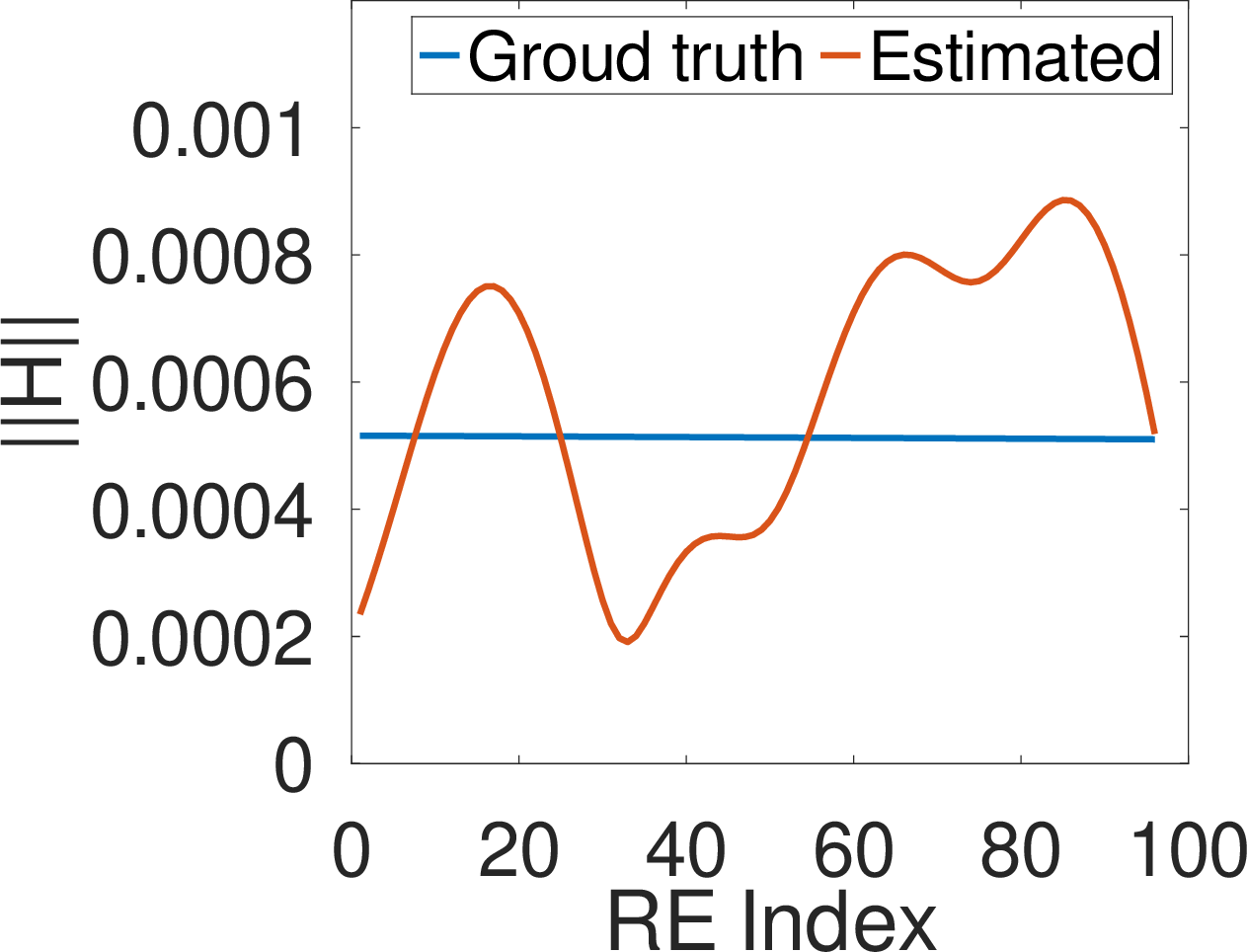}
		\label{fig:ABS}
	}
	\hspace{-0.08in}
	\subfigure[Antenna configure 2] { 
		\includegraphics[width=1.6in]{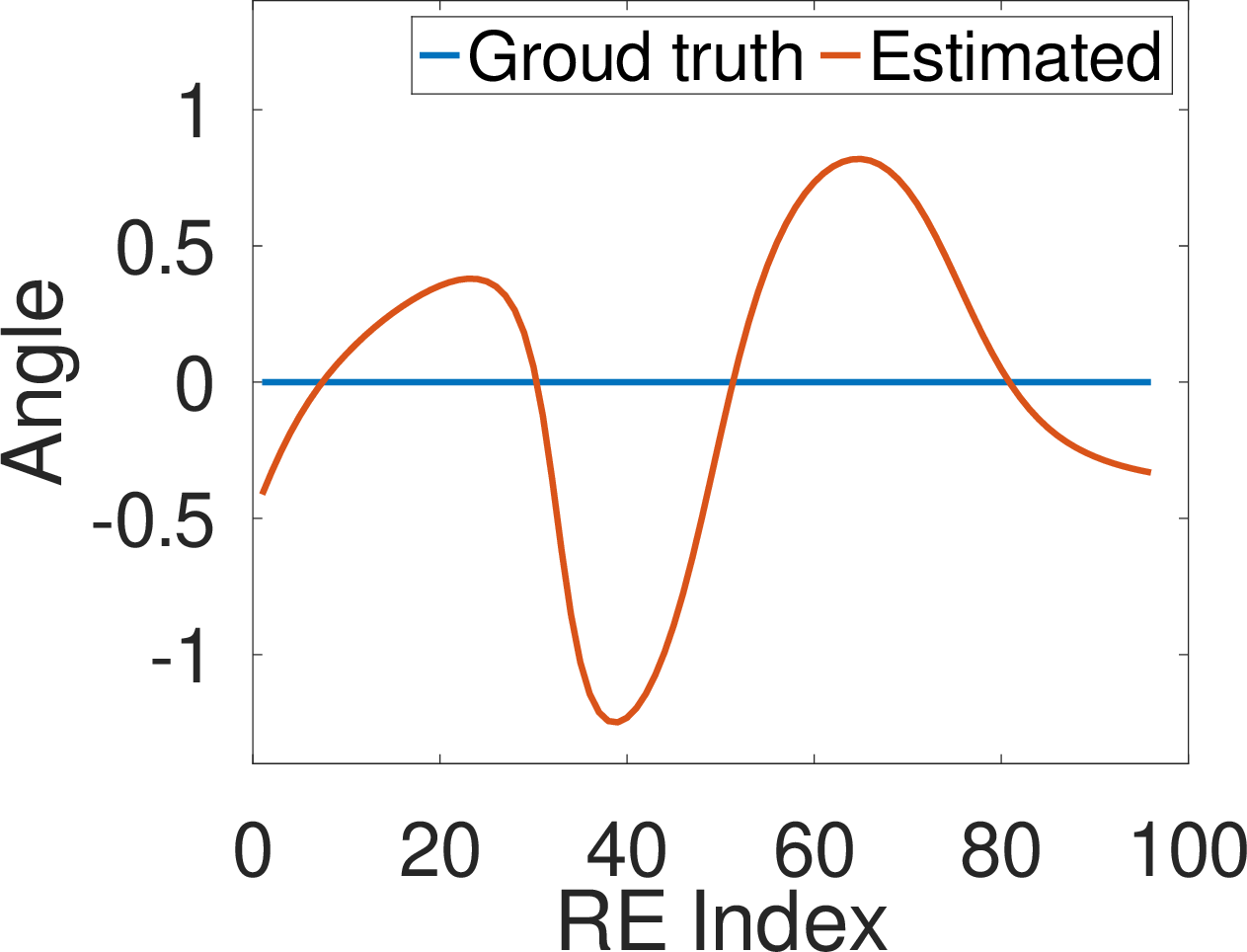}
		\label{fig:ANGLE}
	}
	
	\vspace{-0.05in}
	\caption{Interference Estimation with PHY information}
	\vspace{-0.2in}	
	\label{fig:Compare}
\end{figure}
  
\subsection{Channel Estimators}

Current channel estimation algorithms, such as Zero-Forcing (ZF) and Minimum Mean Square Error (MMSE) can be simplified to provide channel estimator $H$ as $H = Y/X$, where $Y$ is the received SRS/DMRS and $X$ is the transmitted SRS/DMRS signals. However, when interference exists, the interfered channel estimator $H_I$ is changed as
\begin{equation}
	H_I = \frac{Y + \sum_{n=1}^{N}I_n + n_0}{X} = H + \frac{\sum_{n=1}^{N}I_n +n_0}{X}
	\label{eq:H_I}
\end{equation}
where $I_n$ is the signal from $nth$ interference source, $n_0$ is the Gaussian white noise. The interference signals distort the phase and amplitude of the linear channel estimator $H$ into non-linear $H_I$, as shown in Figure \ref{fig:Compare}. Thus, the interference can be computed by comparing the difference between the $H$ and $H_I$ from the same channel within the same frame.

However, without changing the PHY design in the current BS and UE, $H$ and $H_I$ cannot be computed within the same frame. To achieve interference estimation with current PHY design, we are motivated to develop a neural network to learn the interference features by comparing the difference between the common features in datasets of $H$ and $H_I$. We assume the interference varies slowly within 1s to guarantee the transferability of interference features between learning and inference.


\subsection{Interference Mitigation with NN}
After acquiring $H$ and $H_I$, the intuitive approach to mitigate the impact of interference is training a monolithic NN to estimate the features of interference, and remove the interference from the interfered signals. However, due to the randomness in interference signals, a monolithic neural network has a major difficulty in performing both capturing the features and removing the interference in real-time. 
\begin{figure}[ht!]
	\centering
	\vspace{-0.1in}
	\includegraphics[height=1.5in]{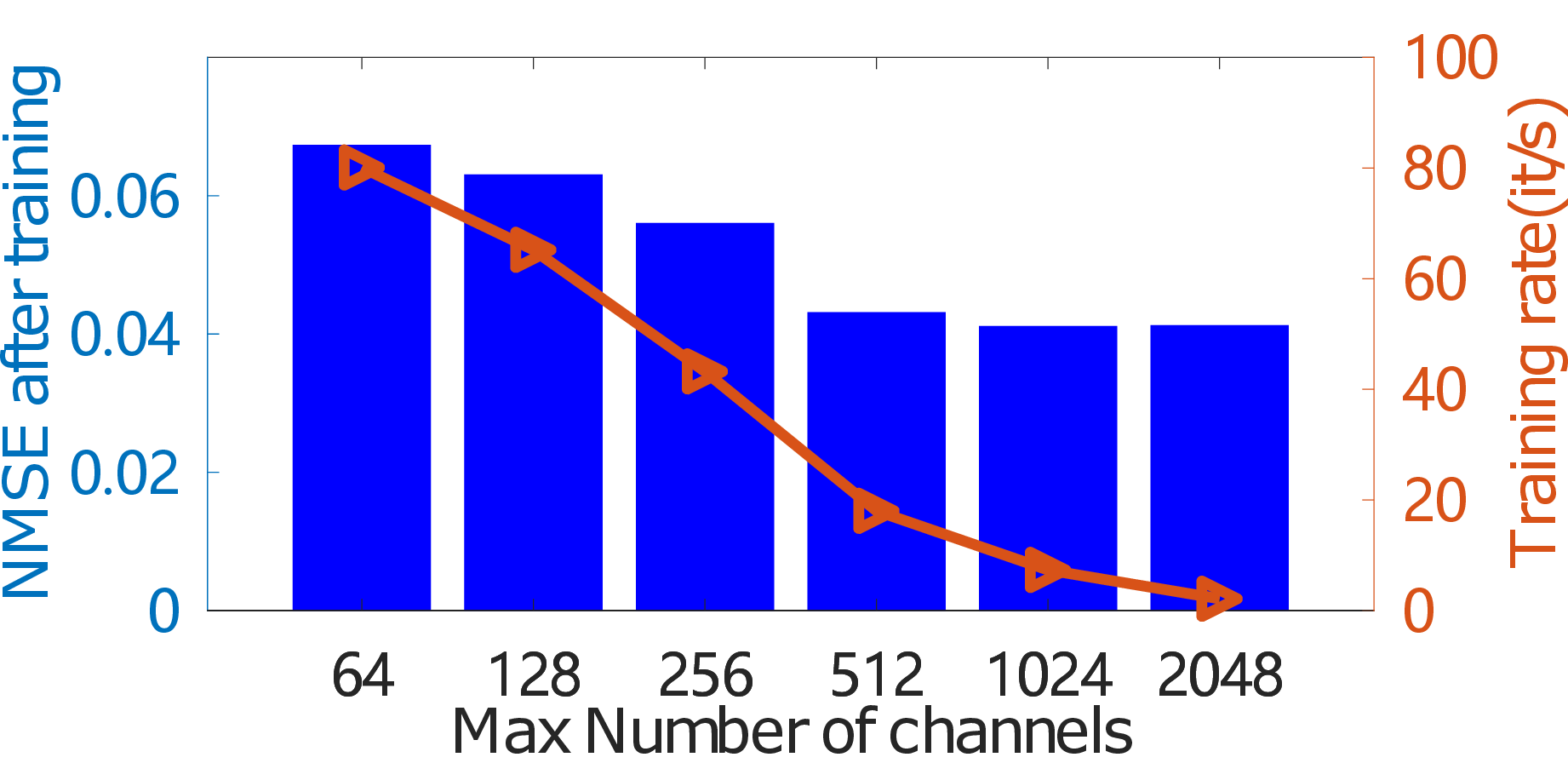}
	\vspace{-0.1in}
	\caption{Interference mitigation with monolithic NN}
	\vspace{-0.1in}	
	\label{fig:Monolithic NN}
\end{figure}

\begin{figure*}
	\centering
	\includegraphics[height=1.2in]{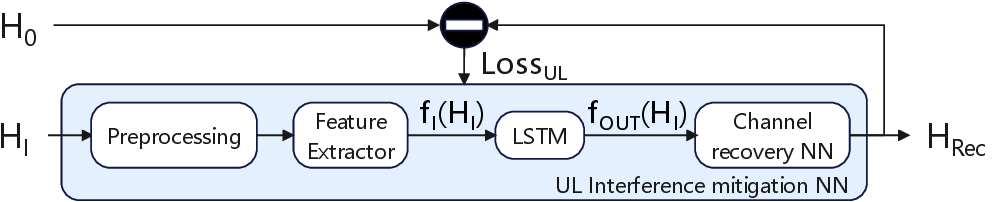}
	\hspace{-0.5in}
	\vspace{-0.1in}	
	\caption{UL NN design}
	\vspace{-0.2in}	
	\label{fig:UL_NN}
\end{figure*}

To verify the effectiveness of a monolithic NN, we conduct an experiment with a 6 layers convolutional NN to mitigate the interference in UL channel estimators, which inputs $H_I$ with NMSE at 0.052 and use $H$ as the training label with 16,32,64 to 512 channels. As shown in Figure \ref{fig:Monolithic NN}, after training, the monolithic NN reduces NMSE for $H_I$ to 0.042 at most, which is only a 20\% reduction due to the confusion in the NN. Based on such insight, we are motivated to develop a neural network that adopts the domain knowledge of PHY in BS and UE to reduce confusion in NN.



\section{Uplink Interference Mitigation}

However, as shown in Figure \ref{fig:Monolithic NN}, training a monolithic NN is costly and inaccurate. To effectively mitigate interference for UL/DL, we design two specialized NNs based on the domain knowledge of UL/DL processing in PHY at BS and UE. 

For UL, BS aims to mitigate the impact of interference in the interfered channel estimator $H_I$ to recover non-interfered channel estimator $H$.  From Eq.\ref{eq:H_I}, we can compute $H$ from $H_I$ and interference as

\begin{equation}
	H = H_I - \frac{\sum_{n=1}^{N}I_n +n_0}{X}
	\label{eq:H}
\end{equation}

Eq.\ref{eq:H} holds when $H$ and $H_I$ share the same physical channel. However, in practice, the frame structure only allows BS to estimate either $H$ or $H_I$ in each frame. Thus, due to the channel fluctuation with UE movement, $H$ and $H_I$ from different frames may not share the same physical channel, causing possible large estimation error in amplitude and phase for interference signals.

\begin{figure}[ht!]
	\centering
	\includegraphics[height=1.2in]{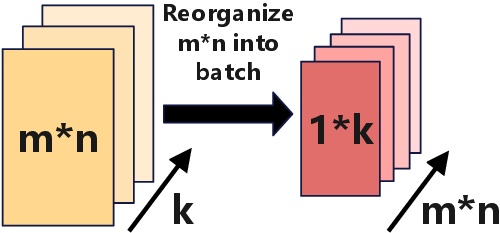}
	\vspace{-0.1in}
	\caption{Preprocessing module}
	\vspace{-0.1in}	
	\label{fig:reorganize}
\end{figure}

Instead, IntLearner aims to exploit the common features from $H$ and compares with real-time $H_I$ input, which allows $H$ to be collected  as a pre-known information for NN training. In the current cellular system, $H$ and $H_I$ are high dimensional data with dimensions of $m*n*k$, where $m$ is the number of BS antenna, $n$ is the number of UE antenna, $k$ is the number of resource elements (RE). The dimension of $H$ and $H_I$ is even larger in Massive MIMO system with $m$ up to 1024 antennas, and $n$ up to 64 antennas. Input such high dimensional data into NN without any pre-processing can lead to a high training cost, due to the greatly increased number of weights in the network.


To reduce such complexity, IntLearner builds a preprocessing module that utilizes the independence between the received channel estimators at each antenna as the domain knowledge to guide the NN training. Due to such independence, IntLearner can focus on training the weights within dimension $k$, and organize the spatial data in $m*n$ into a batch to reduce the total number of weights in NN. Thus, the preprocessing module reshapes $H_I$ from $m*n*k$ to $(m*n)*k$, where $m*n$ is the batch size, as shown in Figure \ref{fig:reorganize}. 

IntLearner further extracts the features $f_{I}(H_I)$ by adding weights to the reorganized $H_I$ as the input as
\begin{equation}
	f_{I}(H_I) = wH_I  
	\label{eq:weights}
\end{equation}

From Eq.\ref{eq:weights}, we design a feature extractor with a 1D convolutional neural network (CNN). Kernel size for the CNN is configured as 2,  which finds the correlations between the channel estimator at neighbor frequency bands based on the short term continuity. However, due to the continuity of interference in the signal bandwidth, long term correlation could exist between distanced frequency bands. To further exploit the long term dependencies within $f_{I}(H_I)$, a LSTM network is implemented to remember the frequency correlation without interference between each RE in $H_I$ by learning the weight in LSTM as
 \begin{equation}
 	 f_{OUT}(H_I) = w_L(wH_I)  
 	\label{eq:lstm_weight}
 \end{equation}

The $f_{OUT}(H_I)$ after LSTM represents the features of $H_I$ after interference mitigation, which should be converted back to the complex number in the frequency domain for later PHY processing. Such conversion is a one-to-one mapping similar to the modulation process in PHY. With the knowledge of modulation process, we designed a channel recovery NN, which compress the features and map it to the corresponding complex value $H_{Rec}$ in the frequency domain.

IntLearner aims to train the optimal weights of $w_L$ and $w$ to compute the $H_{Rec}$ with minimal phase and amplitude distortion to $H$. Based on such knowledge, the loss function for IntLearner in UL is designed to minimizing the difference between $H_{Rec}$ and $H$ available at BS PHY as

\begin{equation}
	Loss_{UL} = min(H_{Rec} - H)  
	\label{eq:Loss_UL}
\end{equation}

\begin{figure*}
	\centering
	\includegraphics[height=1.4in]{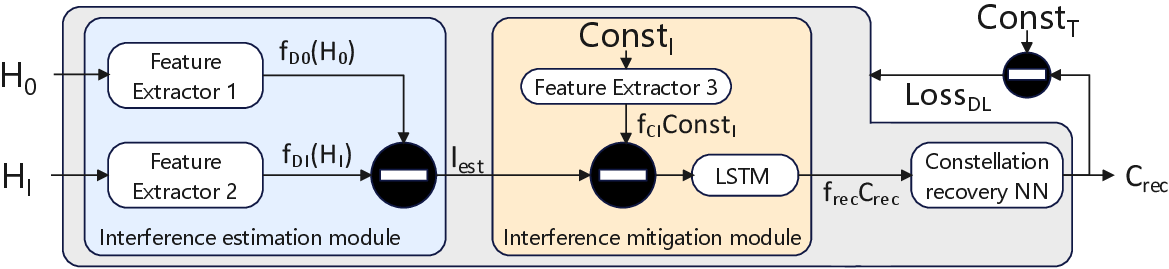}
	\hspace{-0.1in}
	\vspace{-0.1in}
	\caption{DL NN design}
	\vspace{-0.2in}	
	\label{fig:DL_NN}
\end{figure*}





\section{Downlink Interference Mitigation }
UE mitigates the interference in DL by estimating of the interference, and removing such interference from the data symbols. To mimic such process, IntLeaner trains two separated NN modules for interference estimation and mitigation, which takes the domain knowledge of channel estimation, decoding and demodulation as guidance for NN design. 

\subsection{DL Interference Estimation}

From Eq.\ref{eq:H_I}, the interference can be estimated by subtracting $H_I$ by $H$ from the same channel in real-time as

\begin{equation}	
	\sum_{n=1}^{N}I_n  = (H_I - H)X - n_0
	\label{eq:I}
\end{equation}

IntLearner uses Eq.\ref{eq:I} as the domain knowledge to design the interference estimation module to estimate the features of amplitude and phase in the interference signal. IntLearner separately pre-collects two datasets of $H$ and $H_I$, and learns common features from $H$ and $H_I$ by training two sets of parameters $f_{D0}(H)$ and $f_{DI}(H_I)$ with two convolutional NN based feature extractor. During inference, IntLearner reuses the pre-collected dataset of $H$ as a reference to estimate the interference in the real-time input of $H_I$ to reduce the possible errors from applying fixed $f_{D0}(H)$ to different  $H$.



Following Eq.\ref{eq:I},  the interference can be estimated by subtracting $H_I$ by $H$ in the frequency domain. However, IntLearner performs the subtraction in feature space, which may not fully replicate subtraction in the frequency domain. To better mimic Eq.\ref{eq:I} in feature space, we first subtract $f_{D0}(H_I)$ by $f_{DI}(H)$, and add a fully-connected NN to add weights $w_{full}$ for reducing possible estimation error as
\begin{equation}
	I_{est} = w_{full}(f(H_I)-f(H))   
	\label{eq:fully-connected}
\end{equation}

\subsection{DL Interference mitigation}

Before removing $I_{est}$ from the interfered constellation $C_I$ in the interference mitigation module, $C_I$ should be converted to feature space. IntLearner trains a fully-connected NN for adding NN weights $f_{CI}$ to $C_I$ , and outputs the $C_I$ in features space as $f_{CI}(C_I)$.

With $f_{CI}(C_I)$ and $I_{est}$, the interference mitigation NN can mitigate the impact of interference and acquire the recovered constellation $f_{rec}(C_{rec})$ by removing $I_{est}$ from $f_{CI}(C_I)$ as
 
\begin{equation}
	f_{rec}(C_{rec})= f_{CI}(C_I) - I_{est}  
	\label{eq:Const}
\end{equation}

However, $I_{est}$ is estimated from the common features within the channel estimators, which may not precisely measure the interference in real-time. Thus, the impact of interference possibly not be completely removed from the $f_{CI}(C_I)$ due to the limited accuracy in $I_{est}$, which could result in more decoding errors for the following process. 

To further correct decoding errors in constellations, we exploit the correlation embedded between constellation from the encoding process. The LDPC encoder encodes incoming data bits with an encoding matrix and further interleave the encoded bits to generate the transmitted constellation $C_T$, which embeds a correlation between constellations. To learn such correlation, we pass the  constellation after removing $I_{est}$ to an LSTM network, which restores the correct  constellation based on remembered the long and short term dependencies between the constellations during encoding.

The interference mitigation module outputs the corrected constellation in the feature space, which should be transformed to the constellation in complex plane for decoding. Such transformation shares similarities with demodulation process in PHY, which are transforming constellation from the complex plane to data bits. To mimic such demodulation process, IntLearner designs a constellation recovery NN to classify the constellation in features space to the constellation in the complex plane. Such data recovery NN compresses the constellation features $f_{rec}C_{rec}$ after LSTM into an array of features that are further extended for  the classification of output $C_{rec}$ in complex plane.

The target for DL NN is to recover the $C_{rec}$ with minimal hamming distance to the transmitted constellation $C_T$ from BS. Thus, we can derive such hamming distance as

\begin{equation}
	Loss_{DL} =  \sqrt{C_{rec}^2 - C_T^2} 
	\label{eq:Const_T}
\end{equation}
 
As shown in Eq.\ref{eq:Const_T}, we set the training loss $Loss_{DL}$ for IntLearner in DL to minimize the distance between received and transmitted constellation for recovering the most transmitted information from the interfered data symbols. 

\begin{figure}[ht!]
	\centering
	\includegraphics[height=1.8in]{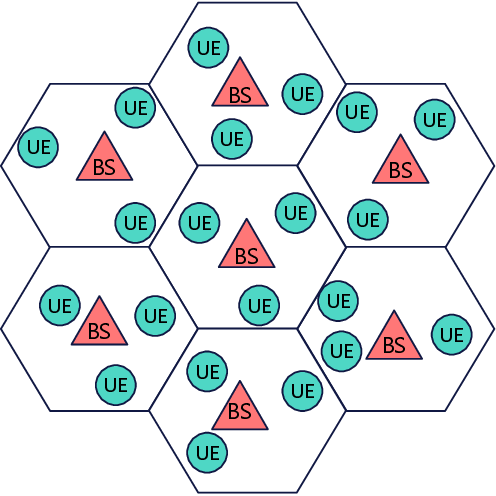}
	\vspace{-0.0in}
	\caption{Multi-cell Simulation Scenario}
	\vspace{-0.1in}	
	\label{fig:sim_setup}
\end{figure}

\begin{figure*}
	\centering
	\hspace{-0.08in}
	\subfigure[Antenna configurations 1] { 
		\includegraphics[width=1.65in]{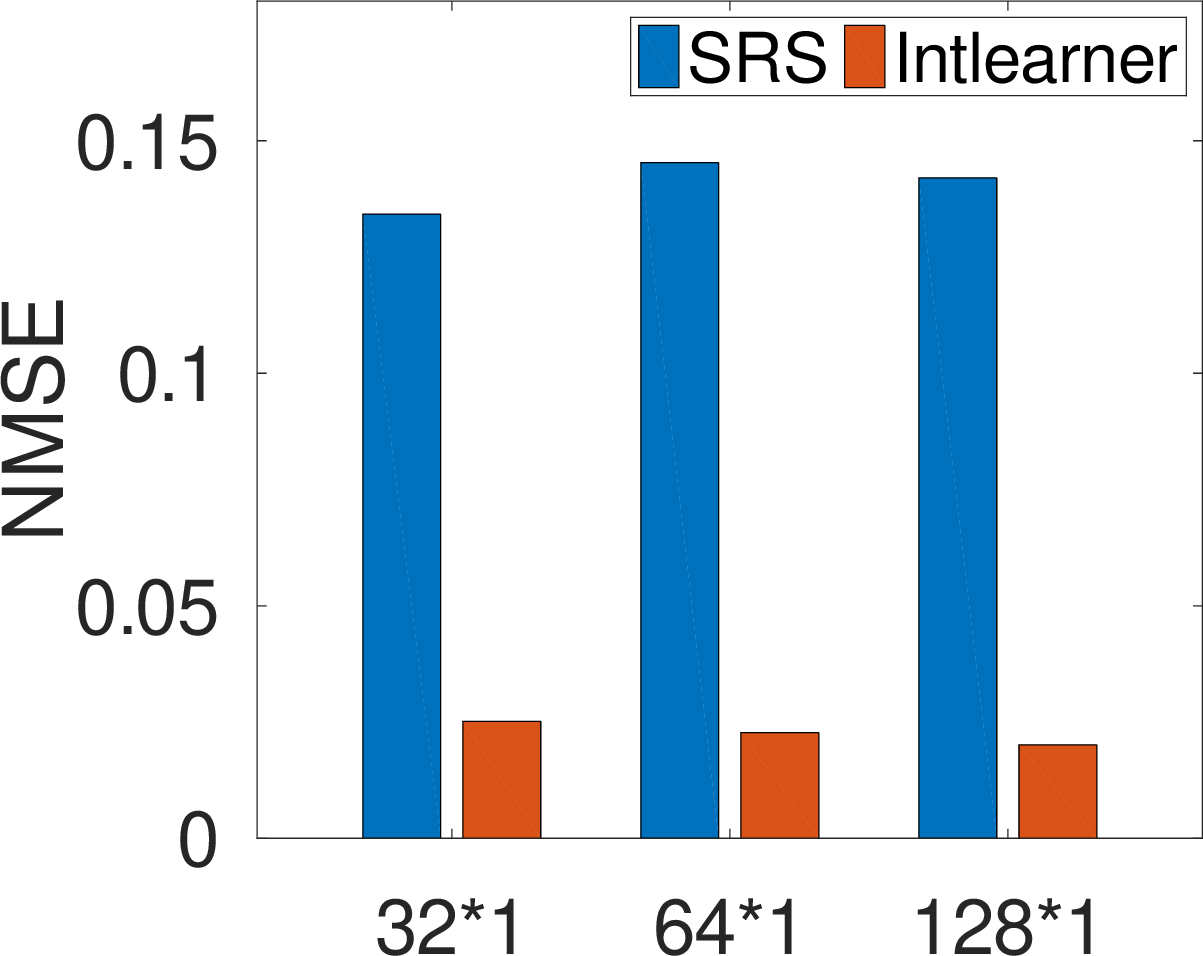}
		
		\label{fig:NMSE_1}
	}
	\hspace{-0.08in}
	\subfigure[Antenna configurations 2] { 
		\includegraphics[width=1.65in]{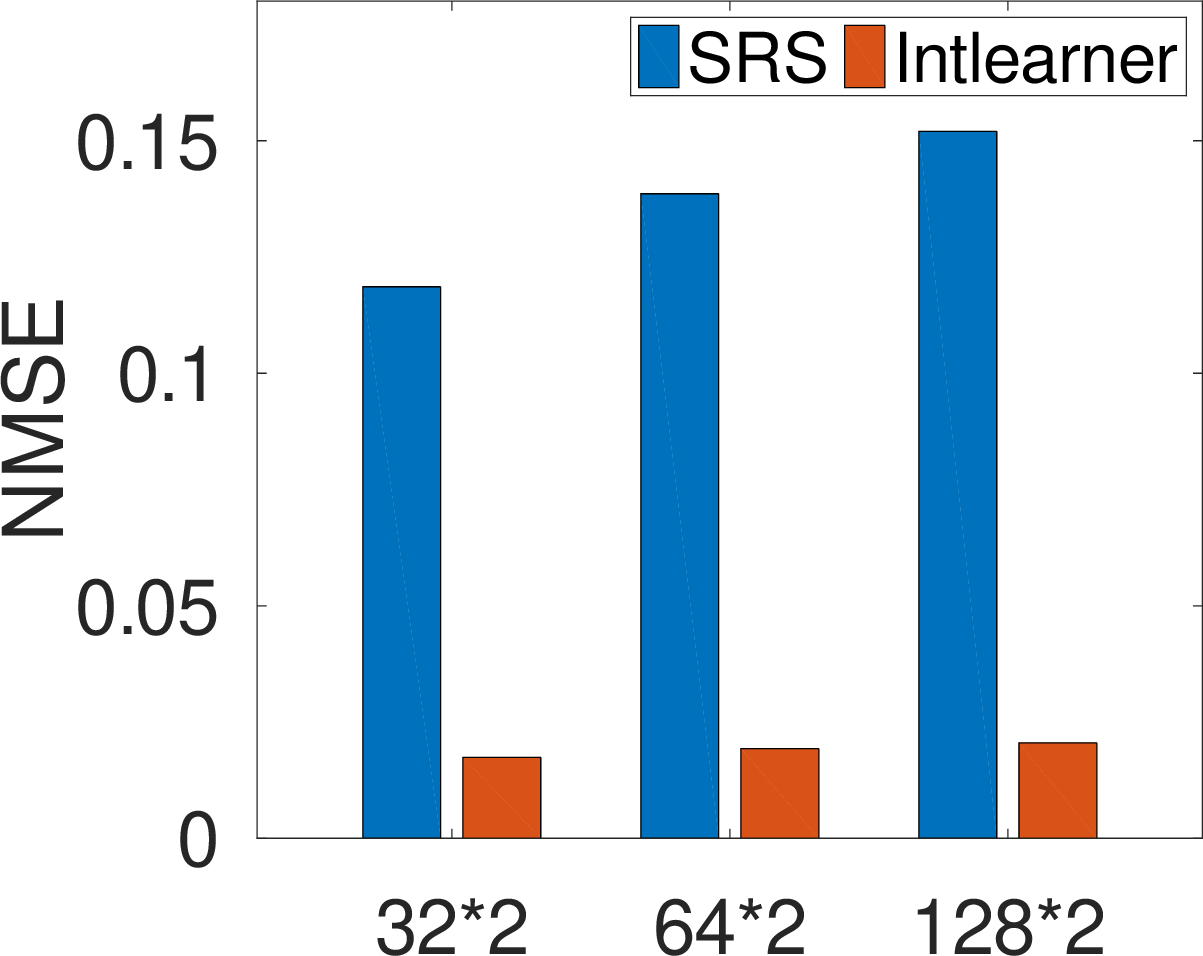}
		\label{fig:NMSE_2}
	}
	\hspace{-0.08in}
	\subfigure[Antenna configurations 3] { 
		\includegraphics[width=1.65in]{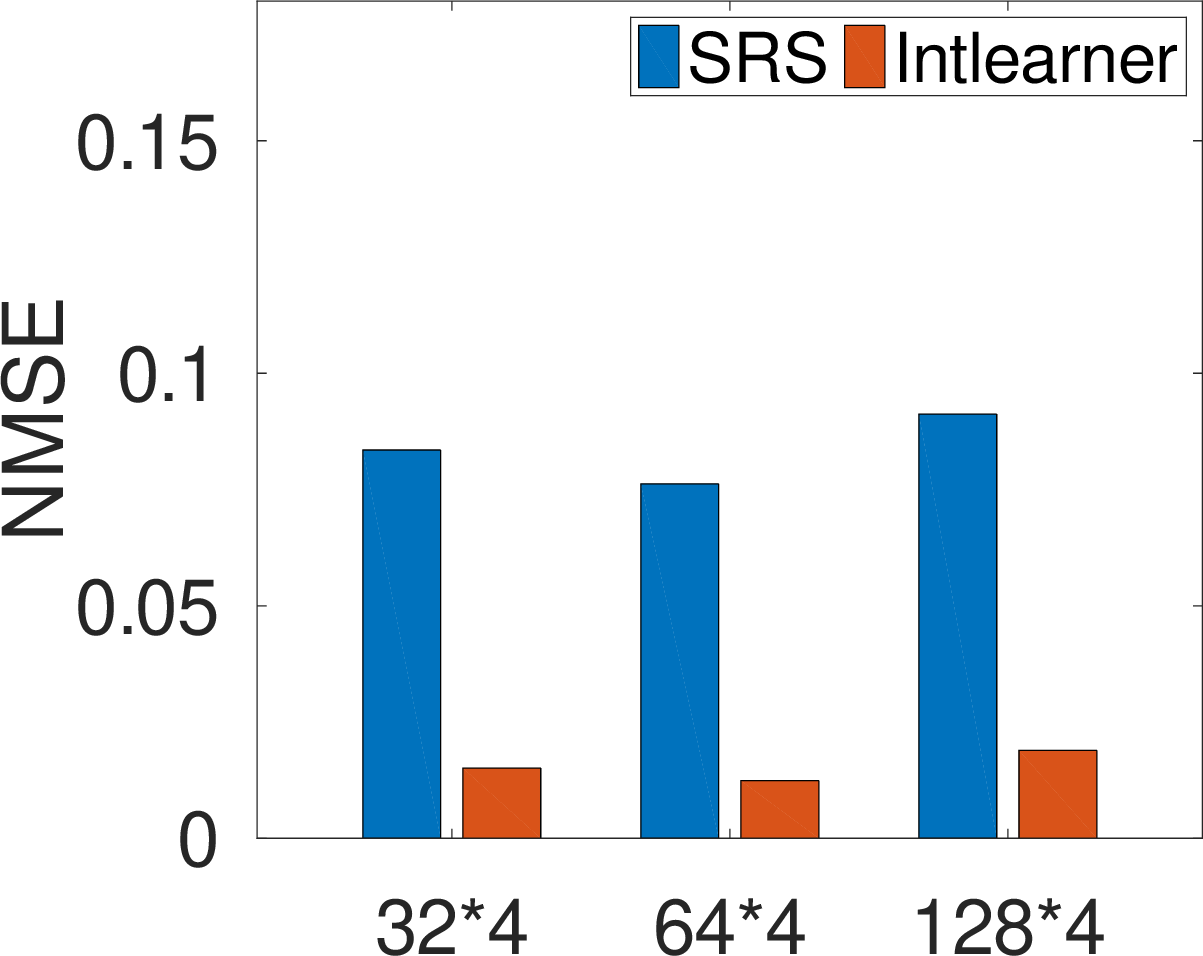}
		\label{fig:NMSE_3}
	}
	\hspace{-0.08in}
	\subfigure[Antenna configurations 4] { 
		\includegraphics[width=1.65in]{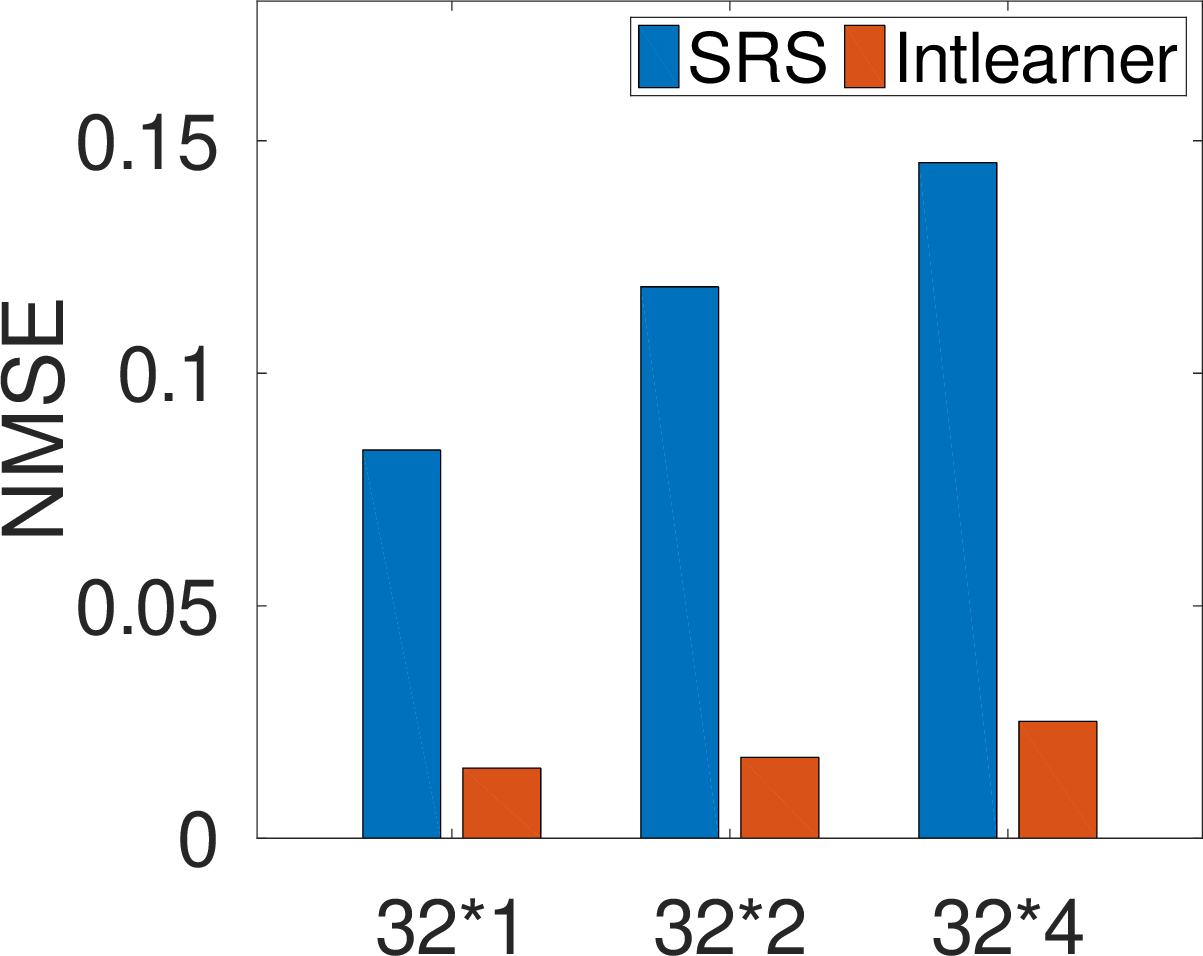}
		\label{fig:NMSE_4}
	}	
	
	\vspace{-0.05in}
	\caption{UL channel estimation NMSE}
	\label{fig:UL_result}
	\vspace{-0.2in}
\end{figure*}

\section{Performance evaluation}

To evaluate the performance of IntLearner, we conduct experiments for mitigating interference in UL and DL, as well as the convergence of NN. For UL, we analyze the Normalize Mean Square Error (NMSE) improvement with IntLearner to process interfered channel estimator. The performance of IntLearner in DL is further studied by comparing the BLER at different SINR and throughput improvement in the center cell. We evaluate the NN complexity and inference cost to verify the practicality of implementing IntLearner in real systems. 


\subsection{Evaluation setup}
We evaluate the performance of IntLearner via a multi-cell and multi-user simulator with random-selected 16 users in each cell.  We configure users in the center cell as the UEs of interest, and the users in six neighbor cells as the interference sources to generate interference signal, as shown in Figure \ref{fig:sim_setup}. All the UEs are randomly located within each cell.  The frequency reuse factor is set at 1:1. Each BS is configured to have 32, 64 and 128 antennas, and the UEs have 1 to 4 antennas. The antenna configurations is set with $BSAnt * UEAnt$, for example, antenna configurations $128*4$ is tested under a scenario with 128 BS antennas and 4 UE antennas. The simulation is performed with standard 3GPP UMa TR38.901 channel model from Quadridga to generate the ideal ground truth channel.

\subsection{UL performance evaluation}

We first evaluate the performance of IntLearner in improving the UL channel estimation quality by measuring the NMSE between the received UL channel estimator and ground truth channel after processing the $H_I$ from standard 5G channel estimation from SRS \cite{3gpp2018nr}.  As shown in Figure \ref{fig:UL_result}, IntLearner can increase 4.8x to 7.4x UL channel estimation accuracy by reducing 80.2\% to 86.5\% of the NMSE in the channel estimation with various antenna configurations. Furthermore, as shown in Figure \ref{fig:NMSE_4}, with an increased number of UE antennas, IntLearner can reduce the NMSE to below 0.015 with above 0.15 NMSE after the standard SRS channel estimation, because IntLearner trains the NN for each decoupled MIMO antennas to achieve minimal error in all antennas. Such results shows generality of IntLearner for mitigating the interference for UL under various system configurations.


\subsection{DL performance evaluation}
\begin{figure}
	\centering
	\hspace{-0.1in}
	\subfigure[Various approaches] { 
		\includegraphics[width=1.65in]{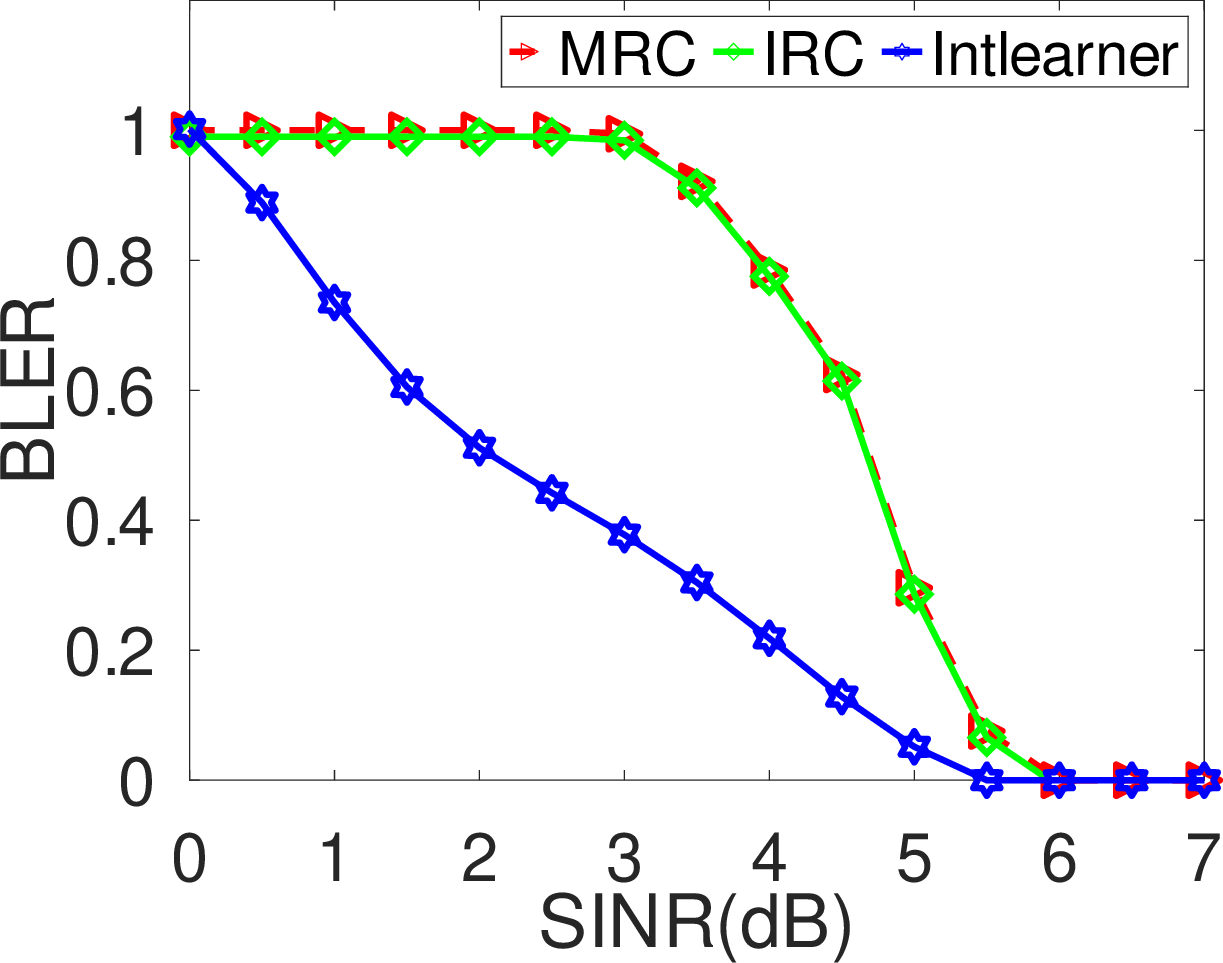}
		\label{fig:BLER_reduction}
	}
	\hspace{-0.1in}
	\subfigure[Different number of antennas] { 
		\includegraphics[width=1.65in]{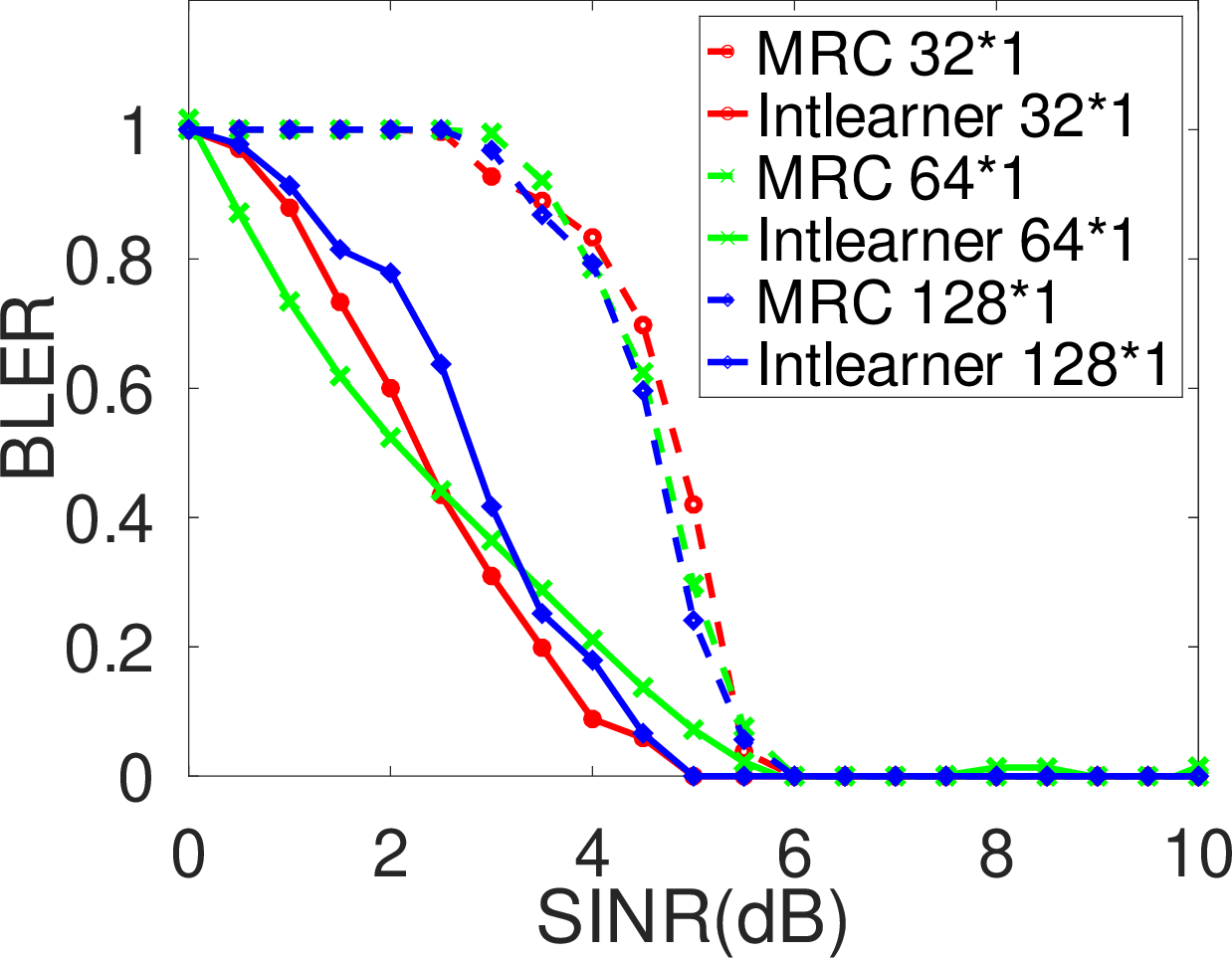}
		\label{fig:Bler_ant}
	}
	
	\vspace{-0.1in}
	\caption{DL BLER reduction}
	\vspace{-0.2in}	
	\label{fig:BLER NN}
\end{figure}

We then evaluate IntLearner's performance in DL by measuring the BLER after receiving 10000 frames with MRC, IRC and IntLearner. The result in Figure \ref{fig:BLER_reduction} shows IntLearner can reduce 1.5dB SINR requirement for each UE to achieve the same 1\% BLER. Moreover, as shown in Figure \ref{fig:Bler_ant}, with different antenna configurations, IntLearner outperforms MRC by reducing 1.5dB SINR requirement to achieve the same BLER, because the NN structure of IntLearner can well adapt to various antenna configurations. Such results demonstrate the generality of IntLearner in mitigating the interference in DL data transmission under different system configurations.


\subsection{NN performance analysis}

\begin{figure}[ht!]
	\centering	
	\includegraphics[height=1.4in]{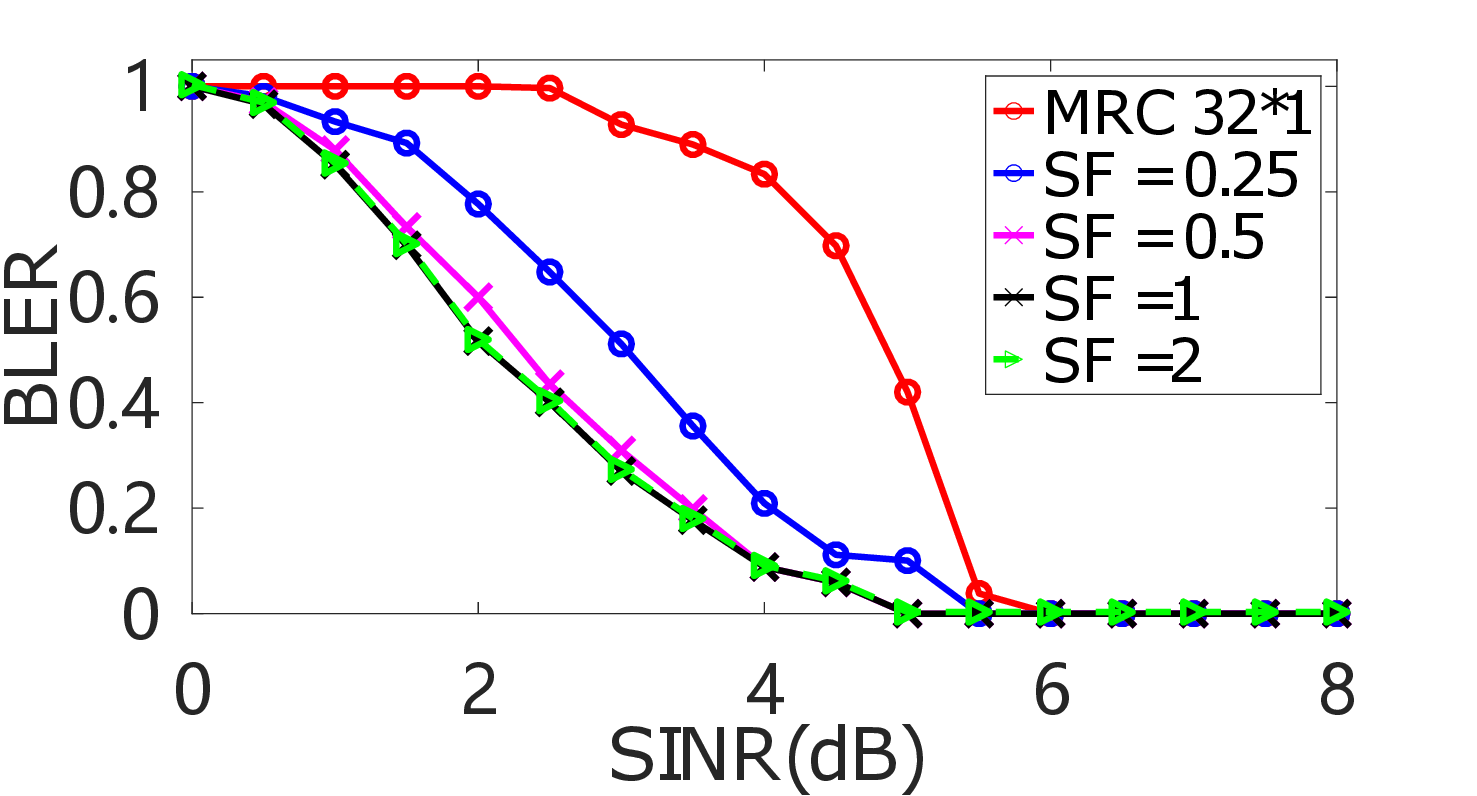}
	\vspace{-0.1in}
	\caption{Mitigating interference with different NN complexity}
	\vspace{-0.2in}	
	\label{fig:NN complexity}
\end{figure}

NN complexity is another key factor that impacts the interference mitigation performance of IntLearner by involving different number of parameters in the neural network. We evaluate NN complexity for IntLearner by fixing the structure of neural work and adding a scaling factor (SF) to the number of nodes at each layers to change the NN complexity. As shown in Figure \ref{fig:NN complexity}, after scaling up the NN complexity, IntLearner can achieve lower BLER at the same SINR. But BLER cannot be further reduced with SF greater than 1, which demonstrates the optimal NN complexity for IntLearner. In practice, the NN complexity can be adjusted to reach target interference mitigation accuracy and hardware limitation.   

\begin{figure}[ht!]
	\centering
	\vspace{-0.1in}
	\hspace{-0.1in}
	\subfigure[Learning curve for IntLearner] { 
		\includegraphics[height=1.25in]{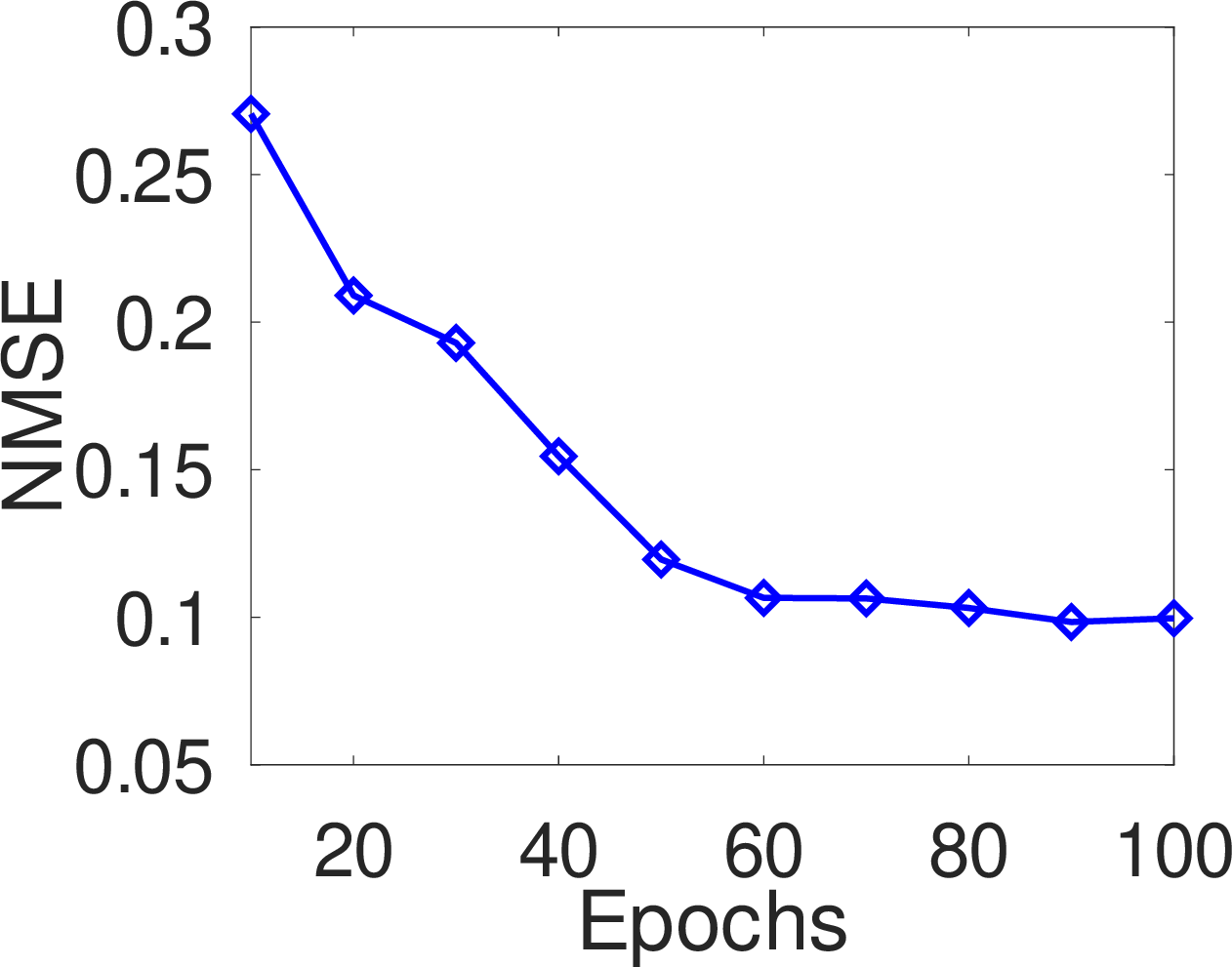}
		\label{fig:learning_curve}
	}
	\hspace{-0.1in}
	\subfigure[Inference time for IntLearner] { 
	\includegraphics[height=1.25in]{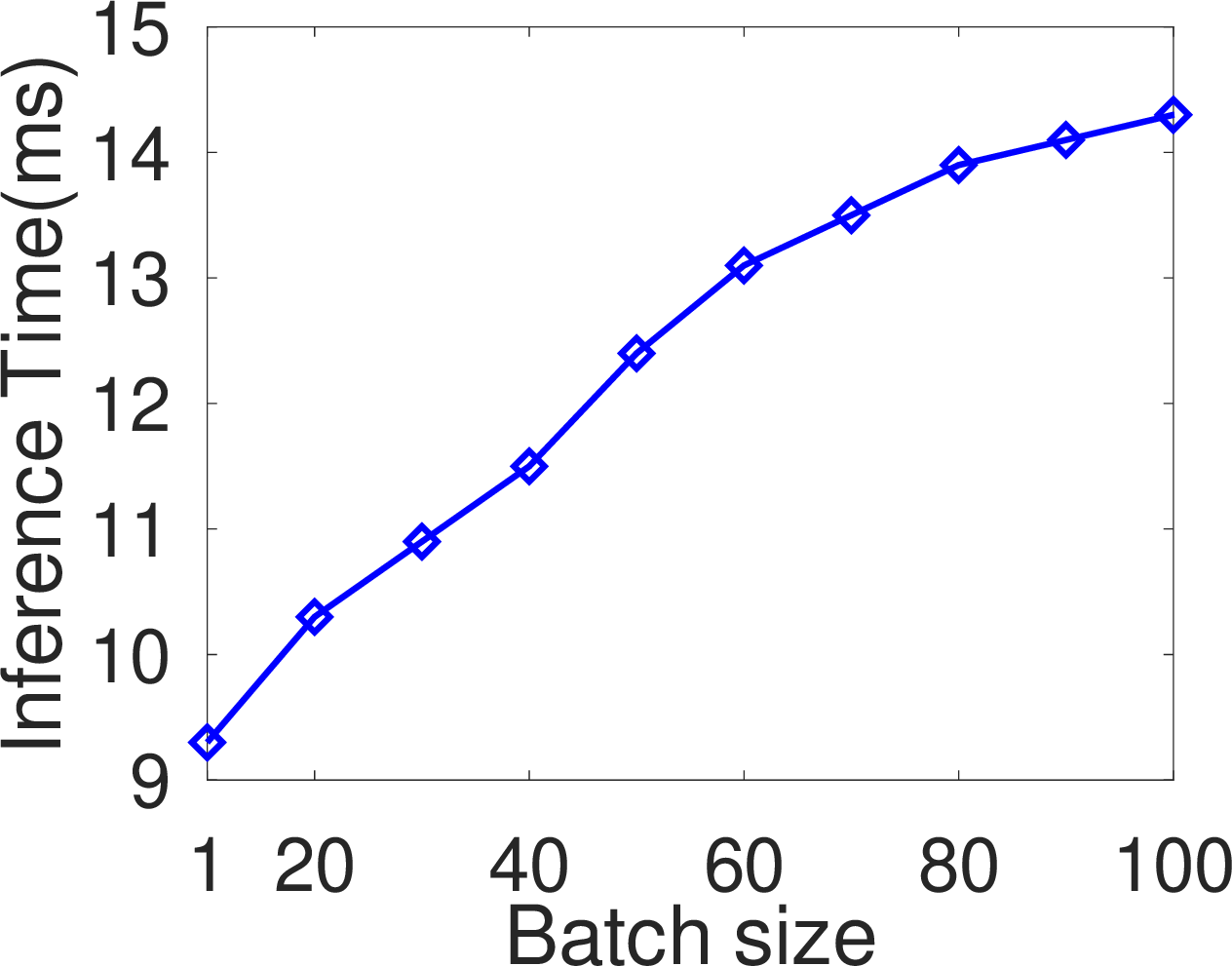}
		\label{fig:Inference_time}
	}
	
	\vspace{-0.05in}
	\caption{Learning and inference time for IntLearner}
	\vspace{-0.1in}	
	\label{fig:Time}
\end{figure}

To implement IntLearner in real cellular system, we expect the NN training to be lightweight and inference cost to be low. To evaluate the cost for training for IntLearner, we evaluate the training curve for IntLearner with various batch sizes. As shown in Figure \ref{fig:learning_curve}, IntLearner NMSE converges after 40 epoch, which is corresponding to 30 minutes of training in our system. The training time can be further reduced by pre-training the model offline and uploading to UE and BS for online refinement after deployment.

We further evaluate the inference time in DL for IntLearner with various batch sizes. As shown in figure \ref{fig:Inference_time}, IntLearner process 100 frames in a batch with about $14.4ms$, the average processing time for each frame is around $144us$, which is only $1/8$ of the $1ms$ frame length in the current 5G system, which indicates IntLearner can achieve real-time interference mitigation in practice.

\section{Conclusion}

In this paper, we proposed IntLearner, a novel AI assisted interference mitigation approach to enhance UL channel estimation accuracy, and reduce DL frame errors from the locally available PHY layer information for next generation cellular networks. IntLearner designs modular NNs guided by domain knowledge of PHY in BS and UE for interference estimation and mitigation without hardware modifications.  With the power of modular NNs, IntLearner enhances the UL channel estimation accuracy up to 7.4x by reducing 86.5\% of NMSE in channel estimation, and reduce the SINR requirement to achieve the same BLER up to 1.5dB in DL communication. 

In the future, we are aiming to extend IntLearner to enhance wireless network performance with interference in DL MIMO systems by designing individual interference mitigation module for each layer. IntLearner can be trained with channels generated with Ray-Traced in various building layouts to better mimic the real-world channel environments. Additionally, the generality of IntLearner in real applications can be further enhanced by training IntLearner with interference data with various interference sources from real-world.

{\small
\bibliographystyle{IEEEtran}
\bibliography{commwlan}

\begin{thebibliography}{10}
\providecommand{\url}[1]{#1}
\csname url@samestyle\endcsname
\providecommand{\newblock}{\relax}
\providecommand{\bibinfo}[2]{#2}
\providecommand{\BIBentrySTDinterwordspacing}{\spaceskip=0pt\relax}
\providecommand{\BIBentryALTinterwordstretchfactor}{4}
\providecommand{\BIBentryALTinterwordspacing}{\spaceskip=\fontdimen2\font plus
\BIBentryALTinterwordstretchfactor\fontdimen3\font minus
  \fontdimen4\font\relax}
\providecommand{\BIBforeignlanguage}[2]{{%
\expandafter\ifx\csname l@#1\endcsname\relax
\typeout{** WARNING: IEEEtran.bst: No hyphenation pattern has been}%
\typeout{** loaded for the language `#1'. Using the pattern for}%
\typeout{** the default language instead.}%
\else
\language=\csname l@#1\endcsname
\fi
#2}}
\providecommand{\BIBdecl}{\relax}
\BIBdecl

\bibitem{eMBB2021Dog}
A.~Dogra, R.~K. Jha, and S.~Jain, ``A survey on beyond 5g network with the
  advent of 6g: Architecture and emerging technologies,'' \emph{IEEE Access},
  vol.~9, pp. 67\,512--67\,547, 2021.

\bibitem{MMTC5G}
P.~Popovski, K.~F. Trillingsgaard, O.~Simeone, and G.~Durisi, ``5g wireless
  network slicing for embb, urllc, and mmtc: A communication-theoretic view,''
  \emph{IEEE Access}, vol.~6, pp. 55\,765--55\,779, 2018.

\bibitem{MMTC2021}
A.~Dogra, R.~K. Jha, and S.~Jain, ``A survey on beyond 5g network with the
  advent of 6g: Architecture and emerging technologies,'' \emph{IEEE Access},
  vol.~9, pp. 67\,512--67\,547, 2021.

\bibitem{MRC1999}
T.~Lo, ``Maximum ratio transmission,'' \emph{IEEE Transactions on
  Communications}, vol.~47, no.~10, pp. 1458--1461, 1999.

\bibitem{MRC2003}
F.~Dietrich and W.~Utschick, ``Maximum ratio combining of correlated rayleigh
  fading channels with imperfect channel knowledge,'' \emph{IEEE Communications
  Letters}, vol.~7, no.~9, pp. 419--421, 2003.

\bibitem{MRC2009}
K.~S. Ahn and R.~W. Heath, ``Performance analysis of maximum ratio combining
  with imperfect channel estimation in the presence of cochannel
  interferences,'' \emph{IEEE Transactions on Wireless Communications}, vol.~8,
  no.~3, pp. 1080--1085, 2009.

\bibitem{MRC2014Tan}
R.~Tanbourgi, H.~S. Dhillon, J.~G. Andrews, and F.~K. Jondral, ``Effect of
  spatial interference correlation on the performance of maximum ratio
  combining,'' \emph{IEEE Transactions on Wireless Communications}, vol.~13,
  no.~6, pp. 3307--3316, 2014.

\bibitem{IRCLTE_2012}
K.~Pietikainen, F.~Del~Carpio, H.-L. Maattanen, M.~Lampinen, T.~Koivisto, and
  M.~Enescu, ``System-level performance of interference suppression receivers
  in lte system,'' in \emph{2012 IEEE 75th Vehicular Technology Conference (VTC
  Spring)}, 2012, pp. 1--5.

\bibitem{IRC2011}
Y.~Ohwatari, N.~Miki, T.~Asai, T.~Abe, and H.~Taoka, ``Performance of advanced
  receiver employing interference rejection combining to suppress inter-cell
  interference in lte-advanced downlink,'' in \emph{2011 IEEE Vehicular
  Technology Conference (VTC Fall)}, 2011, pp. 1--7.

\bibitem{IRC2014liu}
F.~Liu, H.~Zhao, and Y.~Tang, ``An eigen domain interference rejection
  combining algorithm for narrowband interference suppression,'' \emph{IEEE
  Communications Letters}, vol.~18, no.~5, pp. 813--816, 2014.

\bibitem{SIC1994Pat}
P.~Patel and J.~Holtzman, ``Analysis of a simple successive interference
  cancellation scheme in a ds/cdma system,'' \emph{IEEE Journal on Selected
  Areas in Communications}, vol.~12, no.~5, pp. 796--807, 1994.

\bibitem{SIC2011li}
P.~Li, R.~C. de~Lamare, and R.~Fa, ``Multiple feedback successive interference
  cancellation detection for multiuser mimo systems,'' \emph{IEEE Transactions
  on Wireless Communications}, vol.~10, no.~8, pp. 2434--2439, 2011.

\bibitem{SIC2015hig}
K.~Higuchi and A.~Benjebbour, ``Non-orthogonal multiple access (noma) with
  successive interference cancellation for future radio access,'' \emph{IEICE
  Transactions on Communications}, vol.~98, no.~3, pp. 403--414, 2015.

\bibitem{SIC2012mir}
N.~I. Miridakis and D.~D. Vergados, ``A survey on the successive interference
  cancellation performance for single-antenna and multiple-antenna ofdm
  systems,'' \emph{IEEE Communications Surveys \& Tutorials}, vol.~15, no.~1,
  pp. 312--335, 2012.

\bibitem{PHYinfo2015poo}
N.~Poosamani and I.~Rhee, ``Towards a practical indoor location matching system
  using 4g lte phy layer information,'' in \emph{2015 IEEE International
  Conference on Pervasive Computing and Communication Workshops (PerCom
  Workshops)}, 2015, pp. 284--287.

\bibitem{ChanEST2011ber}
P.~Bertrand, ``Channel gain estimation from sounding reference signal in lte,''
  in \emph{2011 IEEE 73rd Vehicular Technology Conference (VTC Spring)}, 2011,
  pp. 1--5.

\bibitem{ChanEst2009hou}
X.~Hou, Z.~Zhang, and H.~Kayama, ``Dmrs design and channel estimation for
  lte-advanced mimo uplink,'' in \emph{2009 IEEE 70th Vehicular Technology
  Conference Fall}.\hskip 1em plus 0.5em minus 0.4em\relax IEEE, 2009, pp.
  1--5.

\bibitem{3gpp2018nr}
3GPP, ``Nr; physical channels and modulation,'' \emph{3rd Generation
  Partnership Project (3GPP), Technical Specification (TS) 38.211}, vol.~9,
  2018.

\end{thebibliography}
}

\end{document}